\documentstyle[aps,twocolumn,prl,psfig]{revtex}

\begin{document}

\title{Non-linear conduction in charge-ordered Pr$_{0.63}$ Ca$_{0.37}$ MnO$_{3}$  : Effect of magnetic fields}
\author{Ayan Guha and A.K.Raychaudhuri}
\address{Department of Physics, Indian Institute of Science, 
Bangalore 560 012, India}
\author{A.R.Raju and C.N.R. Rao}
\address{Chemistry and Physics of Materials Unit, Jawaharlal Nehru Center for 
Advanced Scientific Research, Jakkur,P.O., \mbox{Bangalore 560 064,} India}
                  

\twocolumn[\hsize\textwidth\columnwidth\hsize\csname 
@twocolumnfalse\endcsname

\maketitle
\begin{abstract}
Non-linear conduction in a single crystal of charge-ordered Pr$_{0.63}$Ca$_{0.37}$MnO$_{3}$ has been investigated in an applied magnetic field. In zero field, the non-linear conduction, which
starts at T$<$T$_{CO}$ can give rise to a region of negative differential resistance (NDR) which shows up below the N\'eel  temperature. Application of a magnetic field  inhibits the appearance of NDR  and makes the non-linear
conduction  strongly hysteritic on cycling of the bias current. This is most severe in the temperature range where the charge ordered state melts in an applied magnetic field. Our experiment strongly suggests that application of a magnetic field in the CO regime causes a coexistance two phases.
\end{abstract}
\pacs {}
 
]

Rare earth manganites with general chemical formula Re$_{1-x}$Ae$_{x}$MnO$_{3}$ have attracted current interest because of rich variety of phenomena like colossal magnetoresistance (CMR) and charge ordering (CO)~\cite{Tokura1,Raorev}. For certain values of x, close to 0.5, these compounds undergo a first  order charge-ordering (CO) transition where the Mn$^{3+}$ and Mn$^{4+}$ species arrange themselves alternately in the lattice. This transition leading to charge localization occurs on cooling below a temperature T$_{CO}$, refrerred to as the charge ordering temperature. Orbital ordering  also accompanies the charge ordering and  a long range AFM order sets in at lower temperature (T$_N$ $<$ T$_{CO}$).(For some systems, like Nd$_{0.5}$ Sr$_{0.5}$ MnO$_{3}$ T$_N$$=$T$_{CO}$~\cite{Tokura1}.) The system studied by us, Pr$_{1-x}$Ca$_{x}$MnO$_{3}$ (x = 0.37), happens to be an unique charge ordered system in which the charge ordered state remains insulating for all values of x due to its low tolerance factor. It shows both charge and orbital order below T$_{CO}$ = 240K and the AFM order occurs at T$_N$ = 175K.

A fascinating aspect of the charge ordered/orbital ordered state is that the charge ordered insulator (COI) phase is unstable to a number of external perturbations (like magnetic field~\cite{Tokura1}, electric field~\cite{Tokura2,Ponna1},
optical radiation ~\cite{Ogawa1,Fiebig1} etc.) and can be melted by them. An application of the magnetic field leads to a collapse of the charge ordering gap( $\Delta_{CO}$ ) and there is an insulator -metal transition (melting ) of the 
COI state to an FMM state~\cite{Okimoto98,Biswas992}. Optical radiation seems to create conducting filaments which at low temperatures leads to non-linear transport  and even at much lower temperatures a region of negative differential resistance (with V $\sim$ I$^{-n}$ with 0 $<$ n $<$ 1)~\cite{Ogawa1,Fiebig1}. Application of an electric field also gives rise to non-linear conduction, which seems to have a threshold field associated with it and a broad band noise of substantial magnitude with a power spectrum $\sim$ 1/frequency~\cite{Ayan1}.

A schematic of the phase diagram of the material in the H-T plane studied is shown in the inset of figure~1. In a given field H, the COI state melts into the FMM phase at the melting temperature T$_{MH}$ (marked in the figure).  We designate the region T$_{MH}$ $<$ T $<$ T$_{CO}$ with H $\neq$ 0 (shaded in inset)as the "mixed charge ordered" (MCO)region to distinguish it from the COI (H = 0, T $<$ T$_{CO}$) state as well as from the FMM state (H $\neq$ 0, T $<$ T$_{MH}$). We call it MCO region because we will show below that thie region has co-existence of two phases. In this paper we
address the specific question of the nature of non-linear electronic transport in the COI as well as MCO region for T $<$ T$_{MH}$). We have performed experiments along a constant H line. In particular, we observe that in the COI state, transport is strongly non-linear with a negative differential resistance (NDR) developing for T$<$T$_N$ while in the MCO state, the NDR is strongly inhibited.In addition, in the MCO state  there is a large hysteresis in the I-V characteristics seen on cycling of the bias  current. We ascribe this to creation of a coexisting two phase region in the mixed state.(Hysteresis seen in the I-V curve on cycling the bias current is distinct from the hysteresis in resistivity seen on H field cycling in most past experiments)

Our sample was a \mbox{(4 X 2 X 0.3 $mm^3$)} single crystal of \mbox{Pr$_{0.63}$Ca$_{0.37}$MnO$_{3}$} grown by the floating zone technique. Four linear contact pads of Ag-In alloy was soldered on to the sample in linear four-probe configuration 
with separation $\approx$ 0.25 mm. I-V data were taken with current biasing and with a temperature control better than 10mK. To avoid any memory effects, the sample was heated well above T$_{CO}$, before recording the I-V data at each temperature and magnetic field. We have also measured electrical noise by digitizing the voltage across the sample voltage probes. The power spectrum was obtained by fourier transform of the time autocorrelation function of the voltage after subtracting out the mean. The magnetic field measurements were done using cryogen-free superconducting magnet capable
of producing field upto 15T.  In this communication, to limit the scope of data being presented, we report data only for a field of 8T, which in the temparature range of investigation encompasses all the phases that we need to study.

Fig.1 shows the resistivity ($\rho$) as a function of temperature for various magnetic fields. For H = 0, T$_{CO} \approx$ 240K. In a field of 8T, the T$_{CO}\approx$ 210K and  the CO state completely melts to the FMM state at 80K. We identify this temperature as T$_{MH}$ for 8T field. A magnetic field of 12T arrests the formation of CO state and the sample remains metallic at all temperatures. The observed data are in agreement with previous studies done on the Pr$_{1-x}$ Ca$_{x}$ MnO$_{3}$ system~\cite{Okimoto98}. [ Note: Magnetic susceptibility shows a transition near \mbox{T $\approx$240K} T$_{N}$ $\approx$175K.]

\noindent Fig.2 shows the I-V characteristics at different temperatures in zero magnetic field. The I-V characteristics are linear for T $>$ 260K ($>$ T$_{CO}$) and at all \mbox{T $<$ T$_{CO}$}(i.e, in the COI state), non-linear conduction is observed. One can identify a threshold electric field or a current density beyond which the conduction is strongly non-linear as reported earlier in films of Nd$_{0.5}$Ca$_{0.5}$MnO$_3$ ~\cite{Ayan1}. In the inset of fig.2, we show the noise power measured at 6Hz as a function of biasing current I. At the onset of the non-linear conduction the noise increases rather rapidly. The noise has a 1/f character for  0.02 Hz $<$ f $<$ 20 Hz. Appearance of a strong noise component is a characteristics of the melting process.

At \mbox{T $<$ 170K}, a new component gets added to the non-linearity and a region of negative differential resistance ($dV/dI$$<$0) is observed when the bias current I exceeds a current threshold (I$_{TH}$). I$_{TH}$ decreases as T decreases and at T = 85K , I$_{TH}$ $\approx$ 1.2mA corresponding to a current density $\approx$ 0.5 A/cm$^{2}$.In the region of negative differential resistance(NDR), for I $>$ I$_{TH}$, V $\propto$ I$^{-n}$, where 0 $<$ n $<$ 1. The I-V characteristics   are symmetric and exhibits no hysteresis on current cycling except for T close to T${co}$. The observed data are highly reproducible. Interestingly, there are several similarities of NDR state created by current and that created by optical radiation using a laser~\cite{Ogawa1}. We find that the exponent $n$, which is a measure of the NDR ($dlnV/dlnI$  = -$n$), has a strong temperature dependence. In fig.3 we show $n$ as a function of temperature T scaled by T$_{N}$. We find that $n$ $\rightarrow$ 0 as T / T$_{N}$ $\rightarrow$ 1 and for T / T$_{N}$ $\gg $ 1, $n$ $\rightarrow$ constant( $\approx$ 0.5). A likely explanation for occurence of NDR can be that for I $>$ I$_{TH}$ metallic filaments open up. These filaments being of lower resistance  will provide parallel paths of conduction. This extra current path will decrease the voltage drop across the sample for a given current . As the current increases presumably more such channels open up leading up to a further decrease in V. This will manifest itself in the I-V curve with a region of NDR. Non-linear conduction can also occur due to
depinning of CO domains above a threshold applied field as has been reported earlier~\cite{Ayan1}. But in the region of NDR, we assume conduction through metallic filaments to be the dominant mechanism for non-linear conduction.  The formation of the conducting filament is not a breakdown as the I-V curve is reproducible on cycling. A similar mechanism is proposed for optically produced NDR~\cite{Fiebig1}. In an optically melted CO state in the NDR regime, I $\propto$ V$^{-n}$ with n $\approx$ 2/3 at T = 10K~\cite{Fiebig1}.  Given the similarity, it is expected that both have the  same origin.

We have measured the temperature  rise of the sample with respect to the base ($\Delta T$) by attaching directly a 
thermometer to the sample. $\Delta T$ $\leq $20K at the lowest temperature ($\approx$ 80K) and at the highest power dissipation level (0.1W). At 150K, $\Delta < $10K and its is negligible for T $\geq $ 180K. The power dissipation level where the NDR sets in at the lowest temperature leads to a $\Delta T$ $\approx$ 5K. We also investigated  whether the NDR is caused by this heating effect and rule out heating as the cause for the NDR. However, the heating can have some influence on the value of $n$ at the highest measuring current.

It appears that a strong correlation exists between the onset of the NDR
regime and the  magnetic order at T$_N$ . We illustrate this in figure~3 where  we plot the temperature dependence of the intensity of the (0.5.0.5.0) line as obtained from the neutron diffraction~\cite{Kajimoto1} in a sample of Pr$_{0.65}$ Ca$_{0.35}$ MnO$_{3}$. The temperature dependence of the intensity of this line is a measure of the growth of the AFM order below T$_{N}$. It is clearly seen from the figure that not only  $n$ $\rightarrow$ 0 at T = T$_{N}$, it also follows a temperature dependence which closely matches that of the growth of the AFM order as observed through neutron diffraction. There may be two likely reasons for the appearance of NDR below T$_N$. One reason can be that for T $<$ T$_N$ there may be incommensurate to commensurate transition of the CO or orbital order as has been reported in a closely related composition 
Pr$_{0.5}$ Ca$_{0.5}$ MnO$_{3}$~\cite{Mori1}. The incommensuration which is due to disorder in orbital ordering can inhibit formatiom of such conducting filaments as are needed for the NDR. Alternatively, the AFM order in these materials being of pseudo-CE type ~\cite{Cox1}, there is a FM coupling between the planes which contain zig-zag AFM chains. It may be that below T$_N$, this interplane FM coupling enhances the formation of the metallic filaments which can be made up of FMM phases. If the later hypothesis is true we will not see formation of NDR region in CO systems with CE -type AFM order 
which has only AFM coupling between planes.

Figure~4 shows the I-V characteristics at different temperatures in a 8T magnetic field. In the MCO state, the non-linear conduction persists and is qualitatively similar to that found in 
COI. For both \mbox{T $>$ T$_{CO}$}  and \mbox{T $<$ T$_{MH}$} the I-V behaviour is linear, but the two states have $\rho$ differing by two orders of magnitude. ( All the data were taken after field cooling (FC) in H = 8T from \mbox{T $=$ 300K}. At each temperature, the sample was freshly prepared as a FC sample after warming it upto room temperature 
 to avoid any memory effect.) There is a distinct memory of the previously applied field when the field is changed in  the COI state and on warming up beyond T$_{CO}$ one can erase the memory. Experiments were also carried out under zero-field cooled (ZFC) condition. The I-V characteristics for both FC and ZFC are qualitatively similar, and the ZFC curve consistently showing higher V (for a given I) than the corresponding FC curve at each T. Given the limited scope of this paper we donot discuss the details here.

The inset of figure~4 shows a comparison of the I-V curves of the COI and the MCO states at a representative temperature. The COI and the MCO state differ significantly in two aspects, there is no NDR at any temperature in the MCO state at 8T and there is strong hysteresis in the I-V curve in the MCO region.

An important observation of this investigation is that the melting of the CO state either by temperature  \mbox{( T $\approx$ T$_{CO}$ )} or by a magnetic field is accompanied by strong hysteresis behaviour in the I-V curve (i.e, the I-V curves do not follow each other during current ramping up and down the cycle. This is different from the hysteresis seen on H cycling). This can be seen in figures~4 and 5.The hysteresis 
does not depend on the speed with which the current is ramped up and down. Typically in our experiment one cycle is taken over 40 minutes. The area under the hysteresis curve is defined as the hatched region in figure 5.
 We plot the area under the hysteresis curve as a function of T in the inset of fig 5. It is clearly seen that in the COI state, the hysteresis is observed only for \mbox{$\Delta$ T $\approx$ 10K} below T$_{CO}$. In contrast, in the MCO state, hysteresis persists over an extensive temperature range above T$_{MH}$. In the inset (b) of fig. 5, we show an example of the hysteresis in the MCO state as T $\rightarrow$ T$_{MH}$ from above.

The appearance of hysteresis in the I-V curves can be interpreted as due to the coexistance 
of two phases in the MCO region. There is evidence of such coexisting phases in the TEM data taken near T$_{CO}$ in (La,Ca)MnO$_3$ and (Pr,Ca)MnO$_3$ 
systems near the melting transition(T$\approx$T$_{co}$)~\cite{Chen1}.
In this material under study, it has been seen that the ferromagnetic spin correlations persist below T$_{CO}$~\cite{Kajimoto1}. This is stabilized by the applied field and thus can be the nucleus of the FMM phase. In the MCO state these nuclei get stabilized on cooling and then grow as the second phase.Eventually as T$\rightarrow$T$_{MH}$, the MCO state collapses to the FMM phase. Existence of such coexisting phases will prevent the formation of the metallic filaments needed for the occurence of  NDR. As a result the NDR will be strongly suppressed in this region.

To summarize, we have carried out a systematic investigation of non-linear transport in a CO system in a magnetic field. We find that a negative differential resistance region shows up below T$_{N}$ which is inhibited by application of a magnetic field. In the region 
T$_{co}$ $<$ T $<$ T$_{MH}$, application of a magnetic field creates a region of coexisting phases leading to a strong hysteretic I-V curve which become substantial in the temperature regions close to T$_{co}$ and T$_{MH}$.

A.Guha thanks the CSIR Center of Excellence in Chemistry, JNCASR, for financial support.

\newpage
{\centerline {FIGURE CAPTIONS}}            

(1) FIG.~1.Resistivity as a function of T in different magnetic fields for the sample Pr$_{0.63}$Ca$_{0.37}$MnO$_{3}$. The inset shows the schematic phase diagram.
 
(2) FIG.~2. The I-V curves showing non-linear conduction and negative differential resistance. The inset shows the appearance of large noise at the onset of nonlinear conduction.

(3) FIG.~3. Temperature dependence of the exponent n. The line shows the temperature dependence of the  intensity of the (1/2,0/1/2) line obtained from the neutron experiment~\cite{Kajimoto1}.
 
(4) FIG.~4. The non-linear transport in a magnetic field of 8T. The inset shows a comparison of the I-V data at T = 90K in zero field and in H = 8T.

(5) FIG~5. Hysteresis observed in the I-V curves on cycling of the bias current. The inset (a) shows the temperature dependence of the area under hysteresis close to T${co}$. The inset (b) shows the hysteresis in I-V curve in a magnetic field.

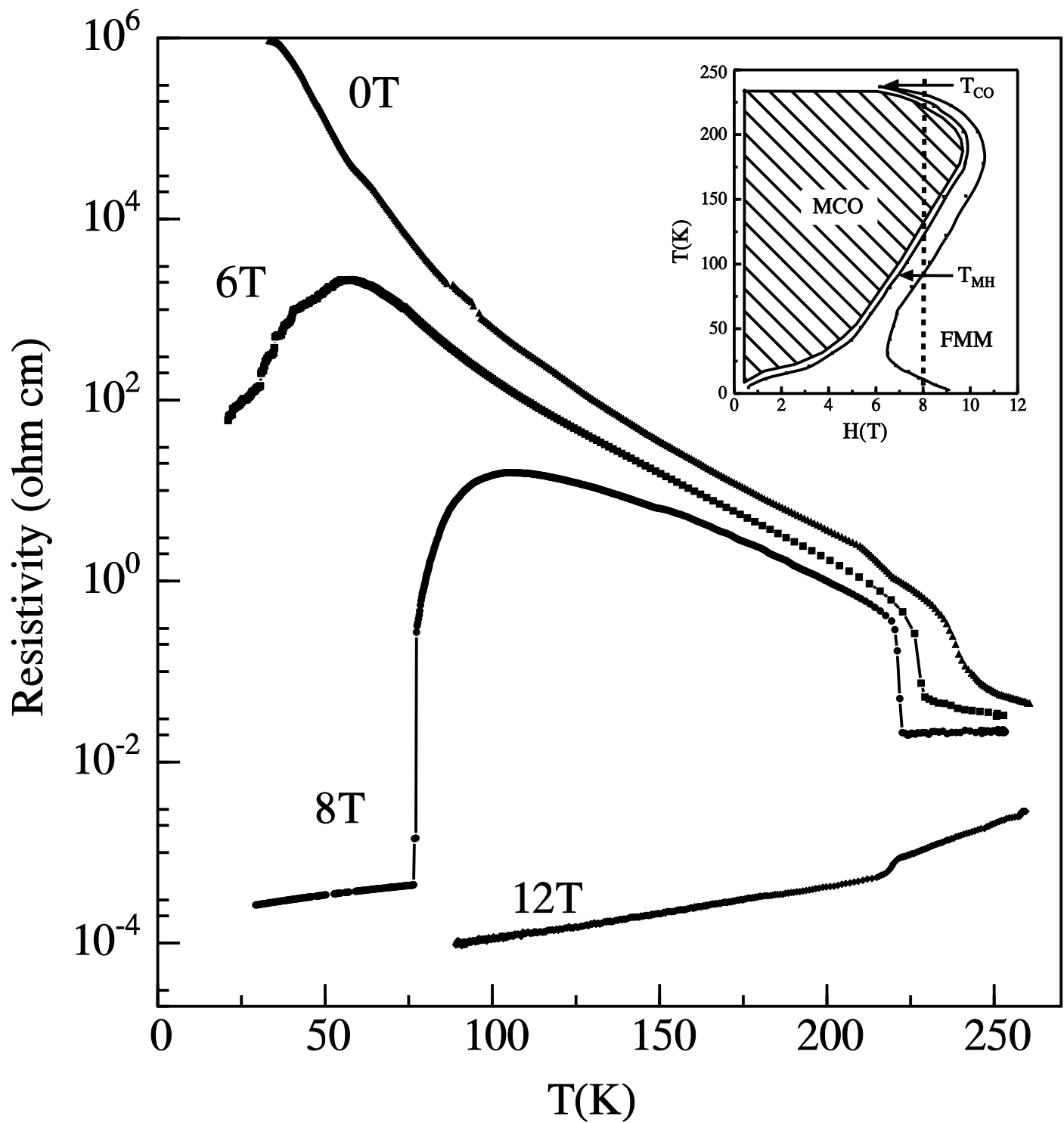
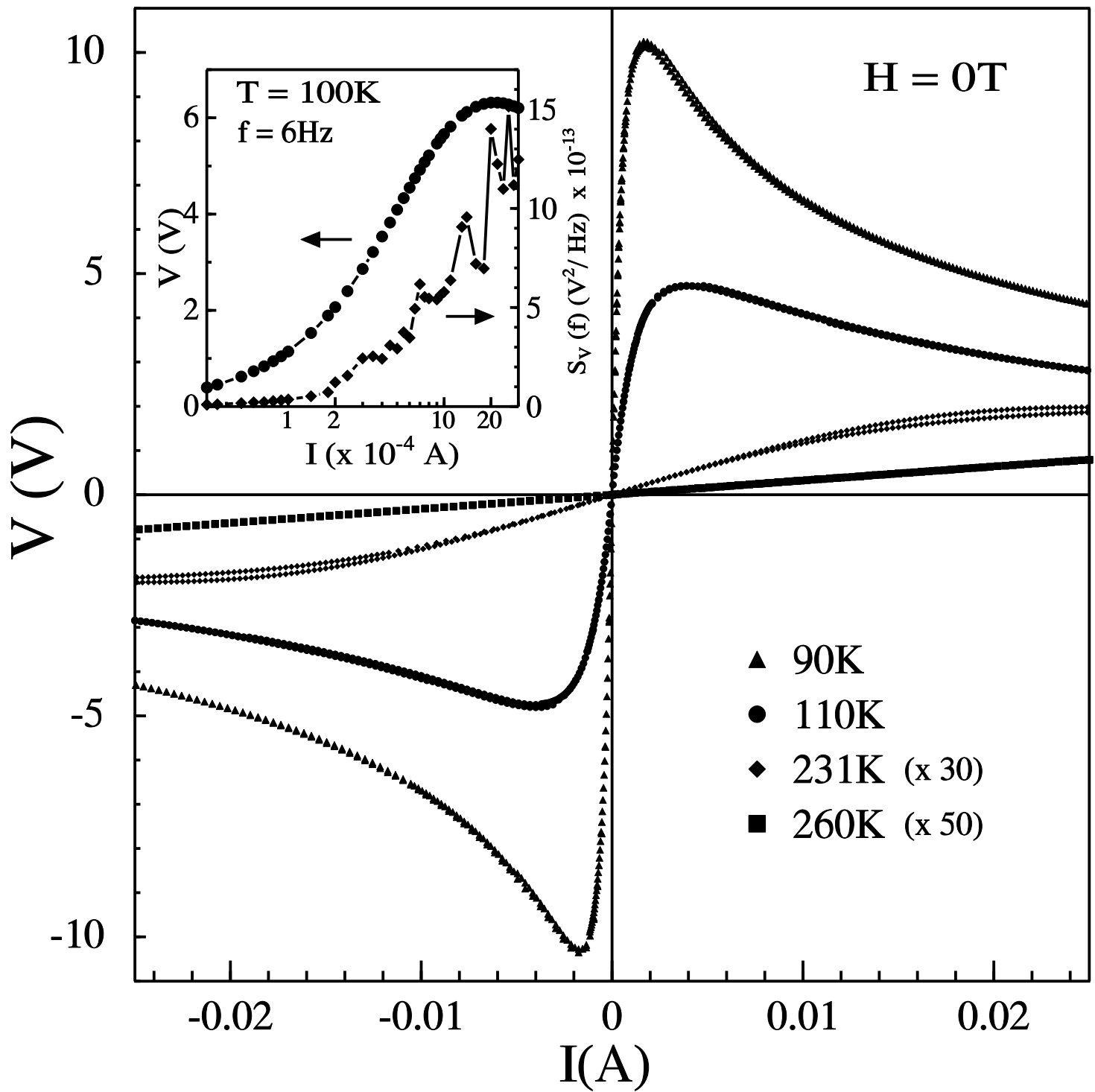
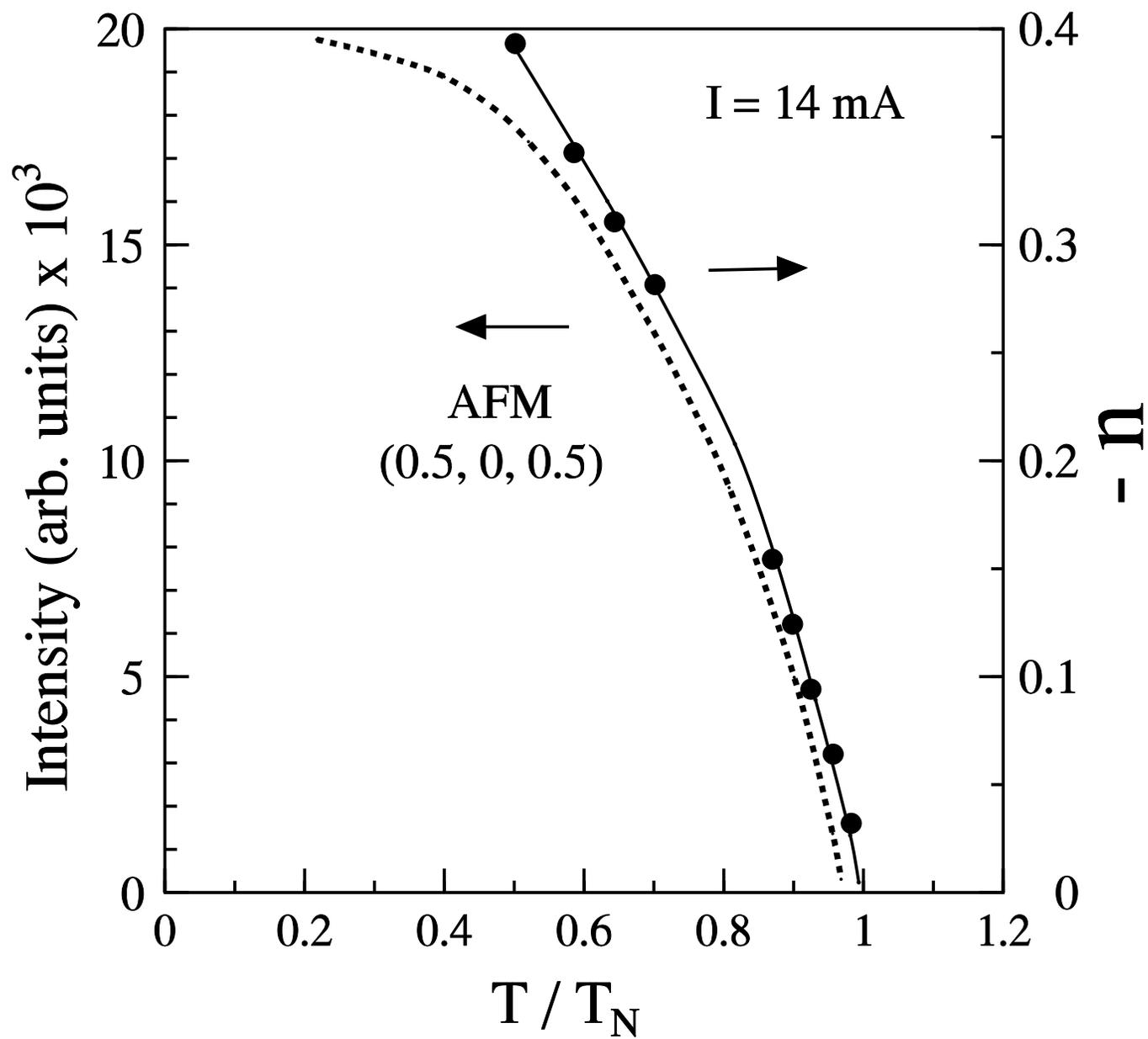
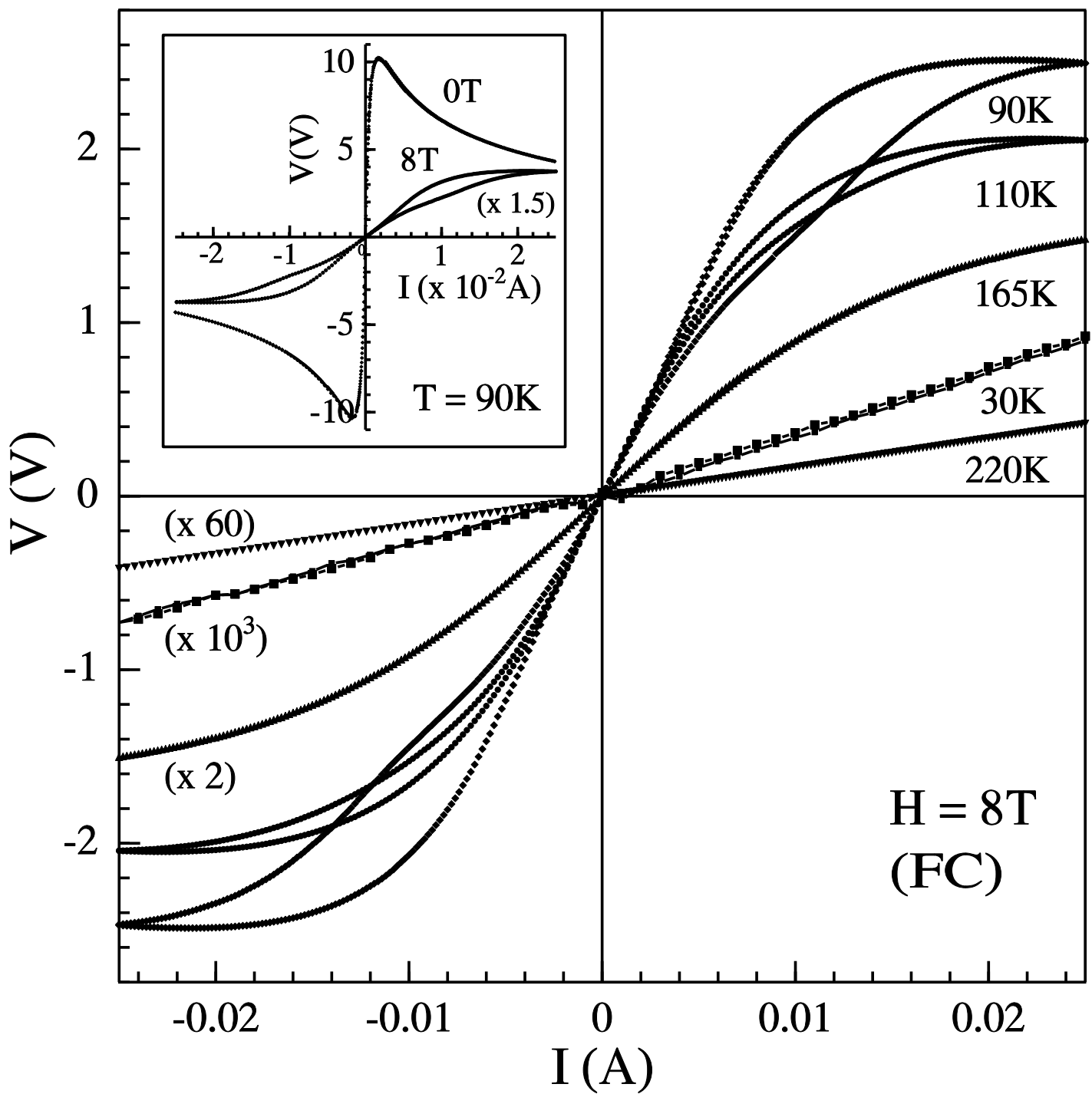
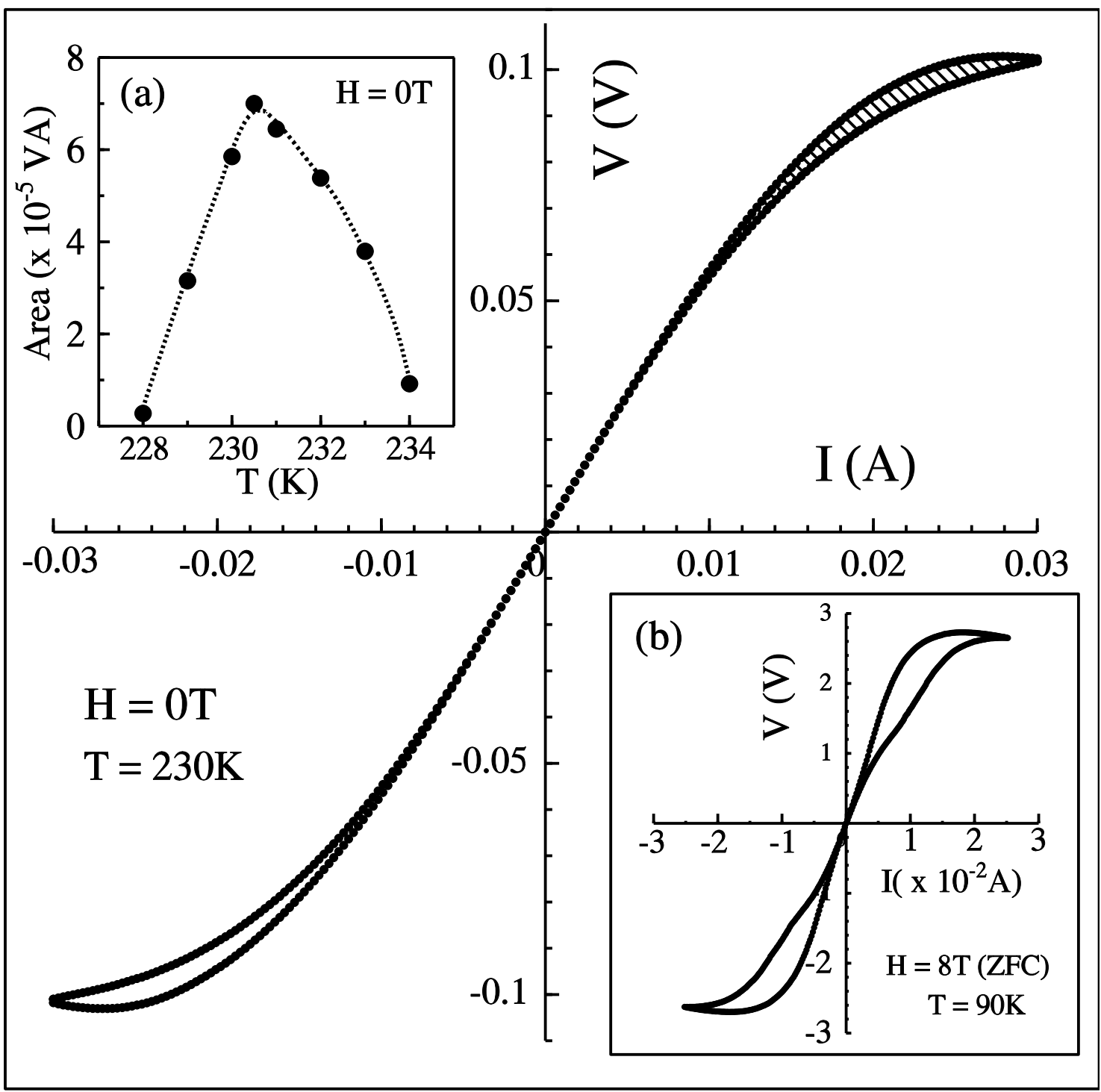
\end{document}